\documentclass[reprint,aip,prl,amsmath,amssymb,reprint,twocolumn]{revtex4-2}

\usepackage{graphicx}                                   
\usepackage{dcolumn}                                    
\usepackage{bm}                                         
\usepackage{chemformula} 
\usepackage[utf8x]{inputenc} 
\usepackage[T1]{fontenc} 
\usepackage{mathrsfs}
\usepackage{braket}
\usepackage{amssymb}
\usepackage{appendix}
\usepackage{hyperref}
\usepackage{color}
\usepackage[normalem]{ulem}
\usepackage{dcolumn}
\usepackage{adjustbox}
\usepackage{tabularx}
\usepackage{tikz}
\usepackage{physics}
\usepackage{program}
\usepackage{graphicx}
\usepackage[normalem]{ulem}
\usetikzlibrary{arrows,shapes,positioning,shadows,trees}

\newcommand{\red}[1]{{\color{black} #1}}

\newcommand{\nl}{\nonumber \\}

\draft 

\begin{document}

\title{From short-sighted to far-sighted: A comparative study of recursive machine learning approaches for open quantum systems} 
\author{Arif Ullah}   
\email{arif@ahu.edu.cn}
\affiliation{School of Physics and Optoelectronic Engineering, Anhui University, Hefei, 230601, Anhui, China}  

\date{\today}

\begin{abstract}
Accurately modeling the dynamics of open quantum systems is critical for advancing quantum technologies, yet traditional methods often struggle with balancing accuracy and efficiency. Machine learning (ML) offers a promising alternative, particularly through recursive models that predict system evolution based on the past history. While these models have shown success in predicting single observables, their effectiveness in more complex tasks, such as forecasting the full reduced density matrix (RDM), remains unclear. 
In this work, we extend history-based recursive ML approaches to complex quantum systems, comparing four physics-informed neural network (PINN) architectures: (i) single-RDM-predicting PINN (SR-PINN), (ii) SR-PINN with simulation parameters (PSR-PINN), (iii) multi-RDMs-predicting PINN (MR-PINN), and (iv) MR-PINN with simulation parameters (PMR-PINN). We apply these models to two representative open quantum systems: the spin-boson (SB) model and the Fenna-Matthews-Olson (FMO) complex. Our results demonstrate that single-RDM-predicting models (SR-PINN and PSR-PINN) are limited by a narrow history window, failing to capture the full complexity of quantum evolution and resulting in unstable long-term predictions, especially in nonlinear and highly correlated dynamics. In contrast, multi-RDMs-predicting models (MR-PINN and PMR-PINN) provide more accurate predictions by extending the forecast horizon, incorporating long-range temporal correlations, and mitigating error propagation. Surprisingly, including  simulation parameters explicitly, such as temperature and reorganization energy, in PSR-PINN and PMR-PINN does not consistently improve accuracy and, in some cases, even reduces performance. This suggests that these parameters are already implicitly encoded in the RDM evolution, making their inclusion redundant and adding unnecessary complexity. These findings highlight the limitations of short-sighted recursive forecasting in complex quantum systems and demonstrate the superior stability and accuracy of far-sighted approaches for long-term predictions.
\end{abstract}
 
\maketitle

Open quantum systems describe quantum systems interacting with their environment, playing a fundamental role in quantum computing, quantum memory, quantum transport, proton tunneling in DNA and energy transfer in photosynthesis.\cite{breuer2016colloquium,khodjasteh2013designing,cui2006quantum,slocombe2022open, zerah2021photosynthetic} Their dynamics are captured by the reduced density matrix (RDM), which evolves under both the system internal dynamics and the influence of its environment.  

Modeling the influence of environment is challenging due to its high-dimensional nature. Mixed quantum-classical methods\cite{Miller2001SCIVR, cotton2013symmetrical, liu2021unified,runeson2019spin,runeson2020generalized, mannouch2020partially, mannouch2020comparison, mannouch2022partially, tao2016multi,mannouch2023mapping, crespo2018recent, qiu2022multilayer} simplify the problem by treating the system quantum mechanically while approximating the environment classically, significantly reducing computational cost. However, these methods often struggle to capture detailed balance\cite{Schmidt2008equilibrium,ellipsoid,thermalization} or subtle quantum correlations.\cite{Mannouch2024coherence} Fully quantum approaches, including path-integral\cite{tanimura1989time, makarov1994path, su2023extended, yan2021efficient, gong2018quantum, xu2022taming, bai2024heom, wang2023simulating, makarov1994path, makri2023quantum} and quantum master equation-based methods,\cite{han2019stochastic, han2020stochastic, ullah2020stochastic, chen2022simulation, dan2022generalized, stockburger2016exact,lyu2023tensor,liu2018exact} provide more accurate descriptions but are computationally expensive, particularly in regimes with strong system-environment coupling or where fine discretization is needed for numerical stability.  

Recently, machine learning (ML) has emerged as a promising tool for learning complex spatiotemporal dynamics in high-dimensional systems.\cite{ullah2025machine, ullah2021speeding, ullah2022predicting, ullah2022one, rodriguez2022comparative, herrera2021convolutional, ge2023four, zhang2023excited, wu2021forecasting, lin2022automatic, bandyopadhyay2018applications, yang2020applications,lin2022trajectory,tang2022fewest, shakiba2024ml_spin_relaxation, lin2024enhancing, zeng2024sophisticated, long2024quantum, cao2024neural, zhang2024nonmarkov, zhang2024ai, herrera2024short,PhysRevResearch.7.L012013} One widely used ML strategy is the recursive approach, where the future evolution of a quantum state is predicted iteratively based on a short history of past evolution. This method has been successfully applied to the relaxation dynamics of the two-state spin-boson (SB) model,\cite{ullah2021speeding, rodriguez2022comparative, herrera2021convolutional, herrera2024short} even enabling extrapolation beyond the trained time window.\cite{ullah2021speeding} However, previous applications have been limited to predicting a single observable—such as the population difference in the SB model—and have relied solely on single-step prediction models.

In this work, we extend recursive ML approaches to more complex quantum systems, focusing on predicting the full RDM rather than just a single observable. We examine four physics-informed neural network (PINN)-based architectures: (i) the single-RDM-predicting PINN (SR-PINN), (ii) the SR-PINN with simulation parameters (PSR-PINN), (iii) the multi-RDMs-predicting PINN (MR-PINN), and (iv) the MR-PINN with simulation parameters (PMR-PINN). These architectures are tested on the relaxation dynamics of the SB model and the exciton energy transfer (EET) process in the Fenna-Matthews-Olson (FMO) complex.

From our results, we underscore the limitations of short-sighted, single-RDM-predicting models (SR-PINN and PSR-PINN) in capturing long-term system dynamics, especially in systems with intricate behavior. These models, constrained by a narrow history window, fail to predict long-term quantum evolution accurately, as they cannot fully capture the complexity of system evolution. In contrast, far-sighted models—such as MR-PINN and PMR-PINN—overcome these limitations by extending the forecast horizon, allowing them to incorporate long-range temporal correlations and achieve more stable predictions.

Although we initially anticipated that incorporating simulation parameters such as reorganization energy ($\lambda$), characteristic frequency ($\gamma$), and temperature ($T$) would improve accuracy, our findings show that these parameters do not consistently enhance performance and, in some instances, actually degrade it. This suggests that the relevant effects of these parameters are already implicitly encoded in the RDM evolution, making their explicit inclusion unnecessary in certain cases.


To build our case, let's consider an open quantum system ($\text{S}$), consisting of \( n \) states interacting with an external environment ($\text{E}$). As stated before, the dynamics of the system is governed by the RDM, which evolves non-unitarily due to environmental effects. While the full system follows unitary evolution described by the Liouville–von Neumann equation, tracing out the environmental degrees of freedom introduces a superoperator $\mathcal{\mathbf{R}}$  that encodes dissipation and decoherence. Under the assumption that the initial state is separable between the system and environment ($\mathbf{\rho}(0) = \mathbf{\rho}_{\rm S}(0) \otimes \mathbf{\rho}_{\rm E}(0) $), mathematically  it can be described as 

\begin{align} \label{eq:neuman}
    \mathbf{\rho}_{\rm S}(t) & = \mathbf{Tr}_{\rm E} \left(\mathbf{U}(t,0) \rho(0) \mathbf{U}^\dagger(t,0) \right) \nl &= -i[\mathbf{H}_{\rm S}, \mathbf{\rho}_{\rm S}(t)] + \mathcal{\mathbf{R}} [\mathbf{\rho}_{\rm S}(t)],
\end{align}
where $\mathbf{\rho}_{\rm S}(t)$ is the RDM of the system at time $t $, $\mathbf{Tr}_{\rm E} $ denotes the partial trace over the environment, $\mathcal{\mathbf{R}}$ is a superoperator that encodes the effects of the environment and $\mathbf{U}(t,0) $ and $\mathbf{U}^\dagger(t,0) $ are the forward and backward time-evolution operators, respectively.

In the recursive ML framework, modeling the time evolution of Eq.~\eqref{eq:neuman} is formulated as learning a mapping function \(\mathcal{M}\) that maps the input descriptors into predicted RDMs. In general, we have
\begin{equation}
    \mathcal{M}: \{\mathbb{R}^{n \times n}\}^{k'} \to \{\mathbb{R}^{n \times n}\}^l \,,
\end{equation}
where \(\{\mathbb{R}^{n \times n}\}^{k'}\) is a collection of \(k'\) input matrices (of size \(n \times n\)) that encode physical information such as historical RDM data, initial conditions, and simulation parameters, and \(\{\mathbb{R}^{n \times n}\}^l\) is a sequence of \(l\) predicted RDMs corresponding to different time steps. In our study, we consider four distinct approaches for predicting the time evolution of RDM:

\hspace{0.5pt}

\noindent\textit{The SR-PINN approach}:  
This method predicts the RDM at the next time step based solely on a fixed-length history of past RDMs. The recursive mapping function is defined as
\begin{equation}
      \mathcal{M}_{\rm rec}: \{\mathbb{R}^{n \times n}\}^{k'} \to \mathbb{R}^{n \times n}\,,
\end{equation}
with
\begin{equation}
     \mathcal{M}_{\rm rec}\Big[\rho_{\rm S}(t_{k-k'+1}),\, \rho_{\rm S}(t_{k-k'+2}),\, \dots,\, \rho_{\rm S}(t_{k})\Big] = \rho_{\rm S}(t_{k+1})\,.
\end{equation}

   The procedure is applied iteratively: after predicting \(\rho_{\rm S}(t_{k+1})\), this new RDM is appended to the history while the oldest entry is removed, keeping the memory size constant at \(k'\).

\hspace{0.5pt}

\noindent\textit{The PSR-PINN approach}:  To improve prediction accuracy, additional simulation parameters \(\mathbf{p}\) (e.g., system–environment coupling, characteristic frequency, temperature) are incorporated into the input. The mapping function becomes
\begin{equation}
     \mathcal{M}_{\rm rec}: \mathbb{R}^p  \times \{\mathbb{R}^{n \times n}\}^{k'} \to \mathbb{R}^{n \times n}\,,
\end{equation}
such that
\begin{equation}
    \mathcal{M}_{\rm rec}\Big[\mathbf{p},\, [\rho_{\rm S}(t_{k-k'+1}),\, \dots,\, \rho_{\rm S}(t_{k})]\Big] = \rho_{\rm S}(t_{k+1})\,.
\end{equation}

   As with the standard SR-PINN, the process is applied recursively with a fixed history length.

\hspace{0.5pt}

\noindent\textit{The MR-PINN approach}:  
   Rather than predicting a single RDM at a time, the MR-PINN approach forecasts a block of future RDMs in one step. Its mapping function is defined by

\begin{equation}
    \mathcal{M}_{\rm rec}: \{\mathbb{R}^{n \times n}\}^{k'} \to \{\mathbb{R}^{n \times n}\}^{N_f}\,,
\end{equation}
with

\begin{align}
       \mathcal{M}_{\rm rec}\Big[\rho_{\rm S}(t_{k-k'+1}),\, \dots,\, \rho_{\rm S}(t_{k})\Big] = \nl \Big[\rho_{\rm S}(t_{k+1}),\, \rho_{\rm S}(t_{k+2}),\, \dots,\, \rho_{\rm S}(t_{k+N_f})\Big]\,.
\end{align}

   In this case, the model outputs $N_f$ future RDMs simultaneously, thus providing a multi-step prediction without requiring iterative updating.

\hspace{0.5pt}

\noindent\textit{The PMR-PINN approach}:  
   This variant extends the MR-PINN method by including simulation parameters in the prediction. The mapping is defined as
\begin{equation}
       \mathcal{M}_{\rm rec}: \mathbb{R}^p \times \{\mathbb{R}^{n \times n}\}^{k'} \to \{\mathbb{R}^{n \times n}\}^{N_f}\,,
\end{equation}
so that

\begin{align}
      \mathcal{M}_{\rm rec}\Big[\mathbf{p},\, [\rho_{\rm S}(t_{k-k'+1}),\, \dots,\, \rho_{\rm S}(t_{k})]\Big] = \nl \Big[\rho_{\rm S}(t_{k+1}),\, \rho_{\rm S}(t_{k+2}),\, \dots,\, \rho_{\rm S}(t_{k+N_f})\Big]\,.
\end{align}

   By integrating the simulation parameters \(\mathbf{p}\), the model can adjust its predictions to account for different physical conditions while forecasting multiple future time steps concurrently.

Each of these approaches leverages the past history of RDMs (and optionally simulation parameters) to predict the future dynamics of the system, differing primarily in whether they predict a single RDM or multiple RDMs in one go.

To evaluate the proposed methods, we analyze the relaxation dynamics of the SB model and the EET process in the FMO complex (see Methods section for details). The models are implemented using a hybrid deep learning architecture that integrates convolutional neural networks (CNNs) with long short-term memory (LSTM) layers, followed by fully connected dense layers (CNN-LSTM). Following the approach outlined in Ref. \citenum{ullah2025machine}, training is optimized using a composite loss function, expressed as:  

\begin{equation}
    \mathcal{L} = \alpha_1\mathcal{L}_1 + \alpha_2\mathcal{L}_2 + \alpha_3\mathcal{L}_3 + \alpha_4\mathcal{L}_4\, ,
\end{equation}  
where each loss term is defined as follows.  

The first term, \(\mathcal{L}_1\), represents the mean squared error (MSE) between the predicted elements of the RDM, \(\mathbf{\rho}_{\text{S}}\), and the reference values, \(\tilde{\mathbf{\rho}}_{\text{S}}\):  

\begin{equation}
  \mathcal{L}_1 = \frac{1}{N_t \cdot n^2} \sum_{t=1}^{N_t} \sum_{i, j = 1}^{n} \left(\tilde{\mathbf{\rho}}_{\text{S}, i,j}(t) - \mathbf{\rho}_{\text{S}, i,j}(t) \right)^2 \, .
\end{equation}  
Here, \(N_t\) denotes the number of time steps.  

To ensure trace conservation of the density matrix, the second loss term, \(\mathcal{L}_2\), penalizes deviations of the trace from unity:  

\begin{equation}
  \mathcal{L}_2 = \frac{1}{N_t} \sum_{t=1}^{N_t} \left( \mathbf{Tr} \, \mathbf{\rho}_{\text{S}}(t) - 1 \right)^2 \, .  
\end{equation}  

The third term, \(\mathcal{L}_3\), enforces positive semi-definiteness by penalizing negative eigenvalues \(\mu_i(t)\) of the density matrix:  

\begin{equation}
  \mathcal{L}_3 = \frac{1}{N_t \cdot n} \sum_{t=1}^{N_t} \sum_{i=1}^{n} \mathbf{\text{max}}(0, -\mu_i(t))^2 \, .  
\end{equation}  

Additionally, \(\mathcal{L}_4\) ensures that all eigenvalues remain within the valid range \([0,1]\), enforcing a key physical constraint of the RDM:  

\begin{equation}
  \mathcal{L}_4 = \frac{1}{N_t \cdot n} \sum_{t=1}^{N_t} \sum_{i=1}^{n} \left(\mathbf{\text{clip}}\left(\mu_i(t), 0, 1\right) - \mu_i(t)\right)^2 \, .
\end{equation}  
The clipping function used here is defined as:  

\begin{equation}
\text{clip}(\mu_i(t), 0, 1) = 
\begin{cases} 
0, & \text{if } \mu_i(t) < 0, \\
\mu_i(t), & \text{if } 0 \leq \mu_i(t) \leq 1, \\
1, & \text{if } \mu_i(t) > 1.
\end{cases}
\end{equation}  

The weighting coefficients \(\alpha_1, \alpha_2, \alpha_3, \alpha_4\) control the relative contributions of these loss terms. In our case, we set them all to unity (\(\alpha_1 = \alpha_2 = \alpha_3 = \alpha_4 = 1.0\)). Collectively, these loss components ensure that the predicted RDM satisfies key physical properties: accuracy (\(\mathcal{L}_1\)), trace conservation (\(\mathcal{L}_2\)), positive semi-definiteness (\(\mathcal{L}_3\)), and eigenvalue constraints (\(\mathcal{L}_4\)).  

For demonstration, we use data from the publicly available QD3SET-1 database\cite{ullah2023qd3set} for both the SB model and the FMO complex. The models are trained on 80\% of the simulations, with the remaining 20\% reserved for testing. Further details on the dataset and training process can be found in the Methods section.

\begin{figure}[!thb]
    \centering
    \includegraphics[width=0.5\textwidth]{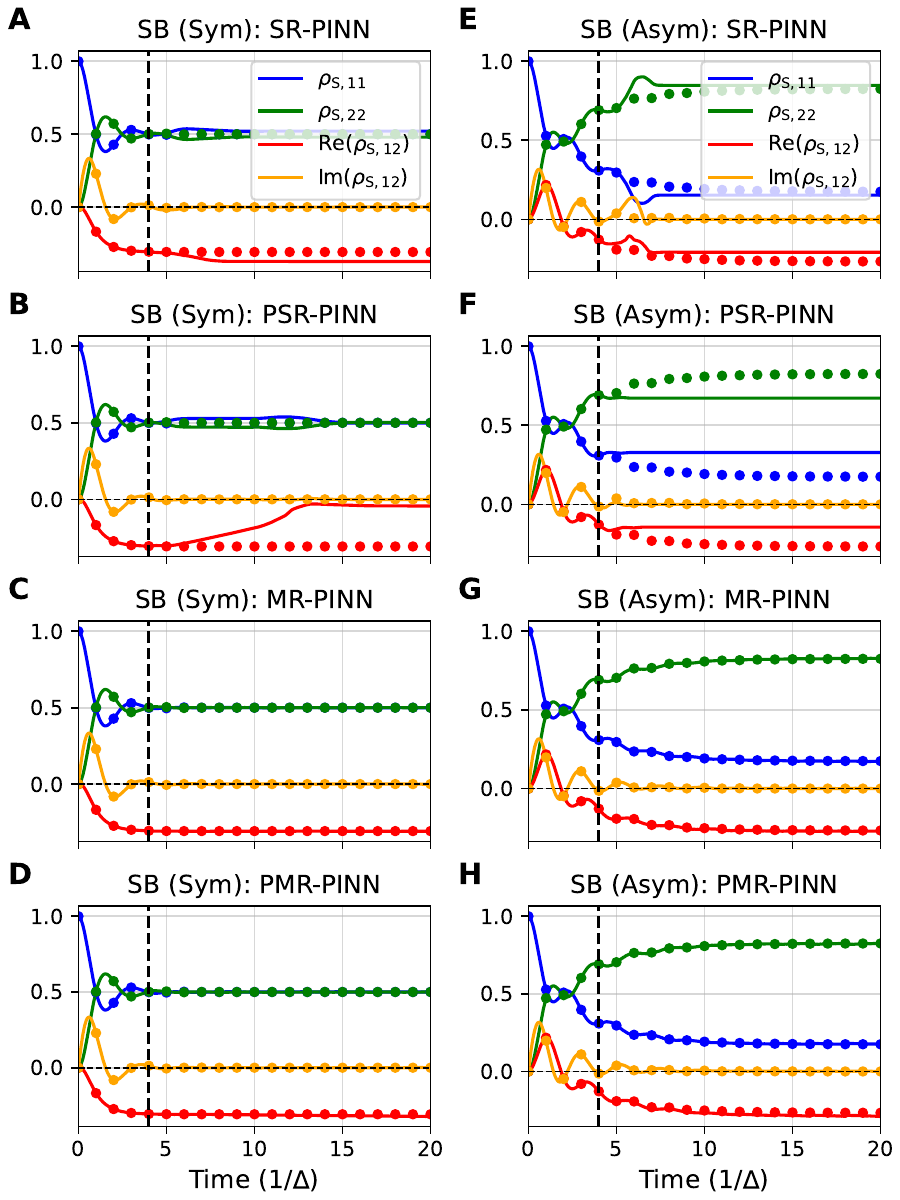}
    \caption{Time evolution of the RDM elements, including both population and coherence terms, as predicted by the SR-PINN, PSR-PINN, MR-PINN, and PMR-PINN models. The first column shows the RDM evolution for the symmetric (Sym) SB model, while the second column displays the corresponding dynamics for the asymmetric (Asym) SB model. Predictions are generated recursively, starting from an initial seed dynamics of time-length \( 4/\Delta \), and are compared to reference results (shown as dots). \red{The dashed vertical line split the seed dynamics (left) and the predicted dynamics (right).} For the symmetric case, the parameters used correspond to an unseen set: \(\varepsilon / \Delta = 0.0\), \(\gamma/\Delta = 3.0\), \(\lambda/\Delta = 0.6\), and \(\beta\Delta = 1.0\). In the asymmetric case, the parameters are \(\varepsilon / \Delta = 1.0\), \(\gamma/\Delta = 9.0\), \(\lambda/\Delta = 0.6\), and \(\beta\Delta = 1.0\).}
    \label{fig:sb_dyn}
\end{figure}

\begin{figure}[!thb]
    \centering
    \includegraphics[width=0.5\textwidth]{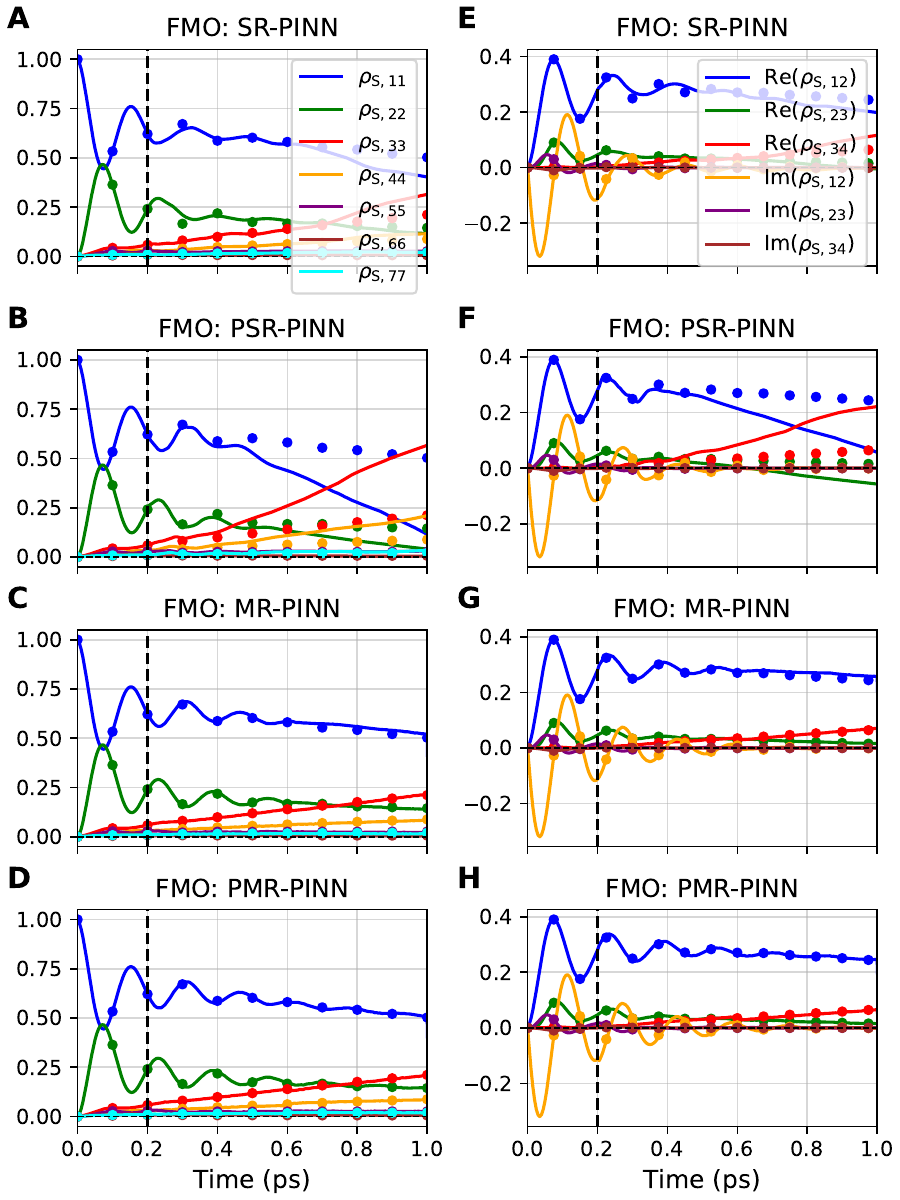}
    \caption{Time evolution of the RDM elements for the FMO complex with initial excitation on site-1, as predicted by SR-PINN, PSR-PINN, MR-PINN, and PMR-PINN. The first column presents the population dynamics of exciton energy transfer (EET), while the second highlights selected coherence elements. Predictions are generated recursively using an initial seed dynamics of 0.2~ps and are compared to reference dynamics (shown as dots). \red{The dashed vertical line split the seed dynamics (left) and the predicted dynamics (right).} The test trajectory corresponds to simulation parameters \(\gamma = 400~\text{cm}^{-1}\), \(\lambda = 40~\text{cm}^{-1}\), and \(T = 90~\text{K}\).}
    \label{fig:fmo_1_dyn}
\end{figure}

Figure~\ref{fig:sb_dyn} presents the predictive performance of all the four models for the time evolution of RDM elements in both symmetric and asymmetric SB models. Each model is provided with an initial short-time seed (\(4/\Delta\)) and tasked with recursively forecasting the system’s future evolution.  

The results highlight the limitations of SR-PINN, which exhibits significant errors in both diagonal and off-diagonal terms (population and coherence), leading to a rapid divergence from the expected dynamics. PSR-PINN, despite incorporating simulation parameters, further degrades accuracy, indicating that the past history window of the model remains insufficient for stable recursive predictions. In contrast, MR-PINN, which leverages a longer forecasting horizon, effectively mitigates error accumulation and successfully captures both population and coherence dynamics across the prediction window. PMR-PINN performs similarly to MR-PINN, suggesting that the inclusion of simulation parameters does not provide additional benefits in this setting.  

To test a larger system, in Figure~\ref{fig:fmo_1_dyn} we showcase the predicted evolution of RDM elements for the FMO complex under initial excitation on site-1. The models are trained with an initial short-time seed (0.2~ps) and recursively predict the system’s future dynamics.  

SR-PINN exhibits considerable inaccuracies, particularly in long-term dynamics, leading to deviations from the expected population transfer trends. PSR-PINN, despite integrating simulation parameters, fails to improve performance and even amplifies errors, especially in diagonal elements. As in the SB model, MR-PINN achieves significantly enhanced accuracy, demonstrating robust predictions of both energy transfer and coherence decay. PMR-PINN yields results comparable to MR-PINN, reinforcing the observation that the longer forecasting window is the primary factor driving predictive stability.  

A quantitative analysis of model performance, summarized in the accompanying table (Table~\ref{tab:mae_results}), further substantiates these findings. As in the SB model, SR-PINN struggles to maintain accuracy, particularly for coherence elements, and PSR-PINN further aggravates errors. MR-PINN consistently outperforms both single-time-step approaches, achieving the lowest mean absolute errors (MAE) across all RDM elements. The inclusion of simulation parameters in PMR-PINN does not lead to meaningful improvements over MR-PINN.  

Notably, the errors are higher for the asymmetric SB model compared to the symmetric case, indicating that prediction accuracy degrades as system complexity increases. This trend suggests that more intricate dynamical behaviors impose additional challenges for PINN-based models, particularly when using single-RDM-predicting training strategies.  

For the FMO complex, the disparity between methods remains evident. SR-PINN and PSR-PINN perform poorly, with PSR-PINN producing the highest errors for both population and coherence terms, particularly when the initial excitation occurs at site-6 (as shown in the table). MR-PINN and PMR-PINN provide a marked improvement, although predictive errors remain higher compared to the SB model, reflecting the increased complexity of the system. The trend observed in the SB model, where errors increase with system complexity, is also evident in the FMO complex. The lack of significant gains from PMR-PINN over MR-PINN suggests that the longer prediction window is the dominant factor in improving accuracy, while the inclusion of simulation parameters has a limited effect.  

\begin{table}[ht]
\scriptsize
\centering
\caption{Time-averaged mean absolute error (MAE) for the diagonal (Diag) and off-diagonal (Off-diag) elements of the RDMs predicted by the SR-PINN, PSR-PINN, MR-PINN, and PMR-PINN models for the test trajectory of the SB model and FMO complex. Off-diagonal errors represent the average MAE for both real and imaginary components. Values are expressed in the form \(10^x\).}
\label{tab:mae_results}
\begin{tabular}{lcccc}
\hline
\textbf{Model} & \multicolumn{2}{c}{\textbf{SB Model (Sym)}} & \multicolumn{2}{c}{\textbf{SB Model (Asym)}} \\
\cline{2-3} \cline{4-5}
               & \textbf{Diag} & \textbf{Off-diag} & \textbf{Diag} & \textbf{Off-diag} \\ 
               &               & \textbf{(Real, Imag)}  &               & \textbf{(Real, Imag)} \\ 
\hline
SR-PINN        & 1.6e-2 & (4.4e-2, 1.7e-3)  & 2.9e-2 & (3.8e-2, 7.8e-3) \\ 
PSR-PINN       & 1.3e-2 & (1.3e-1, 1.1e-3)  & 1.0e-1 & (7.9e-2, 3.6e-3) \\ 
MR-PINN        & 4.9e-4 & (7.3e-4, 5.1e-4) & 1.4e-3 & (3.9e-3, 9.1e-4) \\ 
PMR-PINN       & 6.7e-4 & (5.1e-3, 8.8e-4) & 1.3e-3 & (1.2e-2, 1.2e-3) \\ 
\hline
\textbf{Model} & \multicolumn{2}{c}{\textbf{FMO Complex (site-1)}} & \multicolumn{2}{c}{\textbf{FMO Complex (site-6)}} \\
\cline{2-3} \cline{4-5}
               & \textbf{Diag} & \textbf{Off-diag} & \textbf{Diag} & \textbf{Off-diag} \\ 
               &               & \textbf{(Real, Imag)}  &               & \textbf{(Real, Imag)} \\ 
\hline
SR-PINN        & 1.4e-2 & (4.1e-3, 8.6e-4) & 1.3e-2 & (2.3e-3, 1.6e-4) \\ 
PSR-PINN       & 6.3e-2 & (1.6e-2, 4.8e-4) & 1.1e-1 & (2.9e-2, 2.9e-3) \\ 
MR-PINN        & 2.6e-2 & (6.1e-3, 2.8e-4) & 2.5e-2 & (2.3e-3, 1.5e-4) \\ 
PMR-PINN       & 2.5e-2 & (5.9e-3, 4.2e-4) & 2.6e-2 & (2.3e-3, 1.4e-4) \\ 
\hline
\end{tabular}
\end{table}

In summary, this work investigated four PINN-based architectures for predicting the time evolution of the RDM in selected open quantum systems: single-RDM-predicting models (SR-PINN, PSR-PINN) and multi-RDMs-predicting models (MR-PINN, PMR-PINN). These models use historical RDM data to predict future dynamics, with some incorporating environment-specific parameters such as temperature and system-bath coupling.

Our findings reveal the limitations of single-RDM-predicting models (SR-PINN and PSR-PINN) in capturing long-term quantum dynamics. These models, constrained by a narrow history window, struggle to predict long-term quantum evolution accurately, as they fail to capture the full complexity of system evolution. However, as demonstrated in previous works, \cite{ullah2021speeding, rodriguez2022comparative, herrera2021convolutional, herrera2024short} when trained on simpler observables—such as the population difference in the spin-boson model—they yield reasonable predictions, suggesting that single-variable evolution is easier to propagate recursively.

In contrast, multi-RDMs-predicting models (MR-PINN and PMR-PINN) consistently provide stable and accurate long-term predictions across various scenarios. By predicting multiple RDMs in one step, these models mitigate cumulative errors and better capture long-range temporal correlations, improving their ability to generalize to unseen conditions. This emphasizes that extending the forecast horizon is more effective than merely increasing the historical input length, as explicitly forecasting future states stabilizes predictions more effectively than relying solely on past dynamics.

Surprisingly, explicitly incorporating simulation parameters—such as reorganization energy, characteristic frequency, and temperature (in PSR-PINN and PMR-PINN)—did not consistently improve predictive accuracy and, in some cases, slightly reduced performance. This suggests that the effects of these parameters are already implicitly captured in the RDM evolution, rendering their explicit inclusion redundant and potentially introducing unnecessary complexity.

Overall, this work underscores the limitations of short-sighted, single-step recursive models in complex quantum systems and reinforces the advantages of far-sighted, multi-step approaches for robust, long-term predictions. Our findings highlight that incorporating a longer predictive horizon is key to improving prediction stability, capturing complex dynamics, and reducing the impact of short-term fluctuations in open quantum systems.

\section{Methods}
\noindent \textit{Hamiltonians of the SB model and FMO complex}:  The SB model describes a two-level system interacting with an environment composed of independent harmonic oscillators. The Hamiltonian of the system is given by:  

\begin{equation}
    H = \epsilon \mathbf{\sigma}_z + \Delta \mathbf{\sigma}_x + \sum_{k} \omega_k \mathbf{b}_k^\dagger \mathbf{b}_k + \mathbf{\sigma}_z \sum_{k} c_k (\mathbf{b}_k^\dagger + \mathbf{b}_k),
\end{equation}  
where \(\mathbf{\sigma}_z\) and \(\mathbf{\sigma}_x\) are Pauli matrices, \(\epsilon\) denotes the energy difference between the two states, and \(\Delta\) represents their coupling strength. The surrounding environment consists of harmonic oscillators characterized by creation and annihilation operators \(\mathbf{b}_k^\dagger\) and \(\mathbf{b}_k\), corresponding to mode \(k\) with frequency \(\omega_k\). The system-bath interaction is governed by the coupling coefficient \(c_k\) for each mode.

Our next system of interest, the FMO complex is a trimeric protein found in green sulfur bacteria, where it plays a crucial role in photosynthetic energy transfer. Each monomer of the FMO complex contains multiple chlorophyll molecules—typically seven or eight—that facilitate exciton transport.\cite{am2011eighth} The excitonic dynamics within a monomer can be described by the Frenkel exciton model Hamiltonian:\cite{ishizaki2009unified}   

\begin{align}
    \mathbf{H} &= \sum_{i=1}^{n} \ket{i} \epsilon_i \bra{i} + \sum_{i \neq j}^{n} \ket{i} J_{ij} \bra{j} \nl  
    & + \sum_{i=1}^{j} \sum_{k=1} \left(\frac{1}{2} \mathbf{P}_{k, i}^{2} + \frac{1}{2} \omega_{k, i}^{2} \mathbf{Q}_{k, i}^{2}\right) \mathbf{I} \nl  
    & - \sum_{i=1}^{n} \sum_{k=1} \ket{i} c_{k,i} \mathbf{Q}_{k, i} \bra{i} + \sum_{i=1}^{n} \ket{i} \lambda_{i} \bra{i},
\end{align}  
where \(n\) denotes the number of chlorophyll sites, \(\epsilon_i\) is the site energy, and \(J_{ij}\) represents the electronic coupling between sites \(i\) and \(j\). The operators \(\mathbf{P}_{k,i}\) and \(\mathbf{Q}_{k,i}\) correspond to the momentum and position of the \(k\)-th vibrational mode associated with site \(i\), while \(\omega_{k,i}\) is its frequency. The identity matrix \(\mathbf{I}\) ensures proper dimensional consistency in the model. The coupling strength between site \(i\) and the \(k\)-th vibrational mode is given by \(c_{k,i}\), and \(\lambda_i\) represents the reorganization energy of site \(i\).  

In both the SB model and the FMO complex, the environmental influence is characterized by the Debye spectral density:  

\begin{equation} \label{eq:spectra}
    J(\omega) = 2 \lambda \frac{\gamma \omega}{\omega^2 + \gamma^2},
\end{equation}  
where \(\lambda\) is the reorganization energy, and \(\gamma\) is the characteristic frequency, defined as the inverse of the relaxation time (\(\gamma = 1/\tau\)). For the FMO complex, we assume that all chlorophyll sites experience the same environmental conditions.  

\hspace{0.5pt}

\noindent \textit{Data extraction}: For training our models, we utilized precomputed RDMs provided by the QD3SET-1 database,\cite{ullah2023qd3set} which contains simulations based on the hierarchical equations of motion (HEOM) method.\cite{tanimura1989time, shi2009efficient, chen2022universal, xu2022taming} In the case of the SB model, our dataset, labeled \(\mathcal{D}_{\mathrm{sb}}\), comprises 1000 simulations covering a four-dimensional parameter space defined by $\varepsilon/\Delta$, $\lambda/\Delta$, $\gamma/\Delta$, and $\beta \Delta$, corresponding to the system bias, bath reorganization energy, bath relaxation rate, and inverse temperature, respectively. For the seven-site FMO complex, we also used 1000 simulations from QD3SET-1 that detail the exciton dynamics starting from excitations at site-1 and site-6, spanning the parameter set $(\lambda, \gamma, T)$. In this dataset, the dynamics was generated using the trace-conserving local thermalizing Lindblad master equation (LTLME),\cite{mohseni2008environment} with Hamiltonian parameters taken from the work of Adolphs and Renger.\cite{adolphs2006proteins} Specifically, the FMO Hamiltonian, $\mathbf{H}_{\mathrm{S}}$, is expressed as

\begin{equation}
    \mathbf{H}_{\mathrm{S}} = \begin{pmatrix}
200 & -87.7 & 5.5 & -5.9 & 6.7 & -13.7 & -9.9 \\
-87.7 & 320 & 30.8 & 8.2 & 0.7 & 11.8 & 4.3 \\
5.5 & 30.8 & 0 & -53.5 & -2.2 & -9.6 & 6.0 \\
-5.9 & 8.2 & -53.5 & 110 & -70.7 & -17.0 & -63.6 \\
6.7 & 0.7 & -2.2 & -70.7 & 270 & 81.1 & -1.3 \\
-13.7 & 11.8 & -9.6 & -17.0 & 81.1 & 420 & 39.7 \\
-9.9 & 4.3 & 6.0 & -63.6 & -1.3 & 39.7 & 230
\end{pmatrix},
\end{equation}
with an added diagonal offset of $12210 \, \mathrm{cm}^{-1}$.

\hspace{0.5pt}

\noindent \textit{Data Preparation}: 
To construct the training dataset, each RDM, \(\rho_{\rm S}(t)\), along with its associated coefficients, is flattened into a one-dimensional vector. Given the Hermitian property of the RDM (\(\rho_{\text{S},ij}(t) = \rho_{\text{S},ji}(t)^*\)), we retain only the real components of the diagonal elements while including both the real and imaginary parts of the upper triangular off-diagonal elements.

Each simulation trajectory is then divided into multiple training samples. In the recursive training framework, an initial segment of the system’s dynamics, \(\{\rho_{\rm S}(t_0), \rho_{\rm S}(t_1), \dots, \rho_{\rm S}(t_k)\}\), serves as the input sequence. For the single-RDM-predicting models (SR-PINN and PSR-PINN), the dataset is structured to predict the immediate next RDM, \(\rho_{\rm S}(t_{k+1})\). The input sequence is updated at each step by appending the newly predicted RDM while discarding the earliest one (\(\rho_{\rm S}(t_0)\)), maintaining a fixed sequence length. This iterative process continues until the final time step \(t_K\) is reached.

For the multi-RDMs-predicting models (MR-PINN and PMR-PINN), the output at each prediction step is a block of \(N_f\) future RDMs, predicted simultaneously. The number of predicted steps, \(N_f\), is determined by the length of the prediction window \(t_w\) and the time-step \(dt\) used in propagating the dynamics, via the relation \(N_f = t_w / dt\).

In the SB model, the prediction window is set to \(t_w = 2/\Delta\), leading to \(N_f = 40\) time steps given the chosen \(dt = 0.05/\Delta\). Similarly, for the FMO complex, the prediction window is \(t_w = 0.4\,\text{ps}\), corresponding to \(N_f = 80\) time steps with \(dt = 0.005\)~ps.

More generally, the number of training samples that can be generated from a single simulation depends on the propagation interval \([t_k, t_K]\) and the prediction window \(t_w\), and is approximately given by \((t_K - t_k)/t_w\). These parameters vary depending on the system and the prediction strategy.

For the single-RDM-predicting approaches, a single time-step prediction window is used. In the SB model, we use \(t_k = 4/\Delta\), \(t_K = 20/\Delta\), and \(t_w = dt = 0.05\); for the FMO complex, the settings are \(t_k = 0.2\,\text{ps}\), \(t_K = 1\,\text{ps}\), and \(t_w = dt = 0.005\,\text{ps}\). In contrast, for the multi-RDMs models, the effective \(t_w\) corresponds to the length of the output sequence—\(2/\Delta\) for SB and \(0.4\,\text{ps}\) for FMO—resulting in fewer but richer training samples, each covering a longer prediction horizon.

\red{It is important to emphasize that the selection of \(t_k\) (input sequence length) and \(N_f\) (number of predicted steps) is informed by system complexity, physical intuition, and empirical testing. The value of \(t_k\) should be sufficient to encode the temporal structure and variability of the input RDMs, allowing the model to distinguish between different states. The choice of \(N_f\) depends on the desired prediction horizon and the system's complexity—for instance, \(N_f = 40\) is adequate for the SB model, while the more intricate FMO complex benefits from a longer horizon (\(N_f = 80\)). As a practical guideline, we recommend setting \(N_f\) equal to the time span of the input trajectory to maintain consistency and provide a balanced context for prediction.}

It is important to note that in the PSR-PINN and PMR-PINN models, simulation parameters were normalized by their respective maximum values. The normalized simulation parameters are expressed as \(\lambda/\lambda_{\rm max}\), \(\gamma/\gamma_{\rm max}\), \(\beta/\beta_{\rm max}\) and \(T/T_{\rm max}\), where \(\lambda_{\rm max}\), \(\gamma_{\rm max}\), \(\beta_{\rm max}\) and \(T_{\rm max}\) correspond to the maximum values of \(\lambda\), \(\gamma\), \(\beta\) and \(T\), respectively.  
\hspace{0.5pt}

\noindent \textit{Training and Prediction Strategies}:  
To improve training efficiency, we utilize farthest point sampling\cite{dral2019mlatom, ullah2022predicting} to select a representative subset of simulation trajectories. For each case in SB model (\(\varepsilon/\Delta = 0\) and \(1\)) and the FMO complex (initial excitations on site-1 and site-6), 400 simulation trajectories are allocated for training, with the remaining data reserved for testing. The training is conducted using a CNN-LSTM architecture, where convolutional layers are followed by LSTM layers and fully connected dense layers. For the SB model, a single CNN-LSTM model is trained for both cases (\(\varepsilon/\Delta = 0\) and \(1\)), whereas for the FMO complex, separate models are trained for initial excitations on site-1 and site-6. To ensure a fair comparison, all models share an identical architecture, and during inference, models are selected based on comparable training and validation loss values.

Inference follows the same approach as the training data preparation. In single RDM prediction, a short sequence of past RDMs, \(\{\mathbf{\rho}_{\text{S}}(t_0), \mathbf{\rho}_{\text{S}}(t_1), \dots, \mathbf{\rho}_{\text{S}}(t_k)\}\), serves as the input seed. The model predicts the next RDM, \(\mathbf{\rho}_{\text{S}}(t_{k+1})\), which is then appended to the input sequence while the oldest RDM is removed. This iterative process continues until the entire trajectory is predicted.

For multi-RDMs prediction, a similar strategy is employed, but instead of predicting a single RDM, the model generates a window of \(N_f\) future RDMs in each step. The input sequence is then updated with the last \(N_k\) RDMs from the newly predicted block, allowing for the efficient generation of extended dynamics.

\section{Acknowledgments}

A.U. acknowledges funding from the National Natural Science Foundation of China (No. W2433037) and Natural Science Foundation of Anhui Province (No. 2408085QA002).  

\section{Data availability}
The code and data supporting this work are available at \url{https://github.com/Arif-PhyChem/rc-pinn-comparison}.

\section{Competing interests}
The author declare no competing interests.

\section*{References}
\bibliographystyle{vancouver}
\bibliography{main.bib,references.bib}

\end{document}